\documentclass[aps,prb]{revtex4}

\usepackage{graphicx,amsmath}

\begin{document}
\title{Pseudopotential contributions to the quadrupole moment in charged
periodic systems}

\author{M.J.~Rutter}

\affiliation{TCM, Cavendish Laboratory, JJ Thomson Avenue, Cambridge,
  CB3~0HE, UK}

\email{mjr19@cam.ac.uk}

\begin{abstract} There has been much interest over many years in
  studying charged systems after the artificial imposition of periodic
  boundary conditions, and correcting for the resulting divergence of
  the electrostatic energy density. A correction for cubic cells was
  derived by Makov and Payne in 1995, and its leading error term is of
  the form $L^{-5}$ for a cube of side $L$. Most modern Density
  Functional Theory codes use a different treatment of the `Z$\alpha$'
  energy term to that used by Makov and Payne, resulting in an error
  term of the form $L^{-3}$ if their correction is used unmodified.
  This paper shows how the Makov and Payne result can be made
  consistent with modern practice.
\end{abstract}

\maketitle

\section{Introduction}

Many Density Functional Theory codes assume three dimensional
periodicity which enables the use of simple techniques such as plane
waves for the basis set, and the 3D Ewald sum for electrostatic
interactions. Abinit\cite{Abinit}, \textsc{Castep}\cite{CASTEP},
Quantum Espresso\cite{QE}, \textsc{Vasp}\cite{VASP} and others use
this approach.

If the system to be modelled is not periodic in all three dimensions,
then it can be repeatedly tiled to generate periodicity, with a vacuum
region separating the repeated images. Ideally the calculated
energy, and other quantities of interest, converge rapidly as the size
of the vacuum region is increased.

If the system of interest has zero net charge and no dipole moment,
convergence is reasonably rapid. The worst convergence arises for
systems with net charge, for which the uncorrected energy per unit
volume is divergent. The Ewald sum automatically removes this
divergence by introducing a uniform compensating charge density to
neutralise the cell, but it leaves other terms which are slow
to converge.

The issue was addressed by Makov and Payne\cite{MP95} who
proposed a two-term correction to the energy for cubic cells.
The first term scales as $1/L$ (where $L$ is the cell length), and is
simply the Madelung energy of the lattice. The second scales as
$L^{-3}$ and
depends on the total charge and on the scalar quadrupole moment, $Q$,
(also known as the trace of the quadrupole moment tensor) defined as

\begin{equation}
Q=\int_{\textrm{cell}} r^2 \rho(r) d^3r \label{eqn:quad}
\end{equation}

where $\rho(r)$ is the charge density.

Such corrections, in various geometries, remain the subject of current
research\cite{Freysoldt20,Rutter21,daSilva21}. Historically
corrections for charged systems have been achieved by adding extra
terms to the energy, as Makov and Payne did. They may also be achieved
by adding a constant to the potential so that its average value is no
longer zero\cite{Bruneval14}.

Studies of charged defects in bulk materials need to address further
complications such as the convergence of elastic deformation energies
with cell size, potentially poor localisation of the charge
invalidating expansions based on point multipoles, including the
Madelung energy, and the need to consider the relative permittivity of
the bulk. Such issues produce terms in the energy decaying as slowly,
or more slowly, than the inverse of the volume and prevent even the
$1/L$ energy scaling from being fully
corrected\cite{Freysoldt09,Komsa12}. Furthermore, $Q$ is not
well-defined in periodic systems, being origin-dependent even in the
absence of both a net charge and dipole moment. The value of $Q$ per
unit cell depends on the precise boundary conditions at infinity. This
paper is restricted to isolated charged ions so as to avoid the many
complications introduced by bulk systems.

A pseudopotential differs from the corresponding Coulomb potential
within the pseudopotential's core radius. Thus the integral of the
pseudopotential's potential over all space may differ from that of a
Coulomb potential too. Both integrals are infinite, but their
difference is well--defined. The difference is called the non-Coulomb
$g=0$ term of the pseudopotential, and is usually denoted by
$\alpha$\cite{Ihn79}. It appears in the expression for the total
energy as

\begin{equation}
  E=\frac{|Z|}{V}\sum \alpha_i  \label{eqn:NC}
\end{equation}

where the sum is over all atoms present, $V$ is the cell volume, and
$Z$ is the total charge. Once the system has a net charge, the total
electronic charge and the total ionic charge differ, so it matters
which is used. It had been conventional to use the total electronic
charge\cite{PayneTeter92,Martin04}, but more recent
work\cite{Bruneval14} has shown that the total ionic charge is better
justified, and this later convention is now widely adopted. In some sense
both are correct, for once one tries to reduce an infinite energy, the
energy per unit cell of a 3D-periodic charged system, to a finite
value, the resulting finite value will depend on the conventions used
for the reduction.

\section{A Pseudopotential's `quadrupole moment'}

A dipole, $p$, consisting of two equal and opposite charges of
magnitude $p/r_0$ separated by a distance $r_0$ produces a potential
which extends to infinity. The integral of that potential over all
space is zero if one integrates over the volume of a sphere centred on
the centre of the dipole. This can be seen by considering the symmetry
operation of changing the sign of all the charges, which must change
the sign of the integral, followed by a rotation to make the
sign-changed system identical to the old, an operation which will have
no effect on the integral.

A similar argument applies to a quadrupole moment consisting of
alternating charges arranged on the corners of a square. But a
quadrupole is a tensor, and not all of its components can be
represented thus. The scalar quadrupole moment, $Q$, described by
equation~\ref{eqn:quad}, is represented by a spherical shell of charge
of radius $r_0$ and magnitude $Q/r_0^2$ together with a compensating
central point charge of magnitude $-Q/r_0^2$.

At radii greater than $r_0$ a spherical Gaussian surface contains no net
charge, and, by symmetry, can have no field. Assuming that the potential
is zero at infinity, it is zero at all radii $>r_0$. For radii less than
$r_0$, the point charge produces the usual $1/r$ potential, and the
spherical shell a constant potential given by the the value for a
point charge at $r_0$. One can determine the integral of the total
potential given by this model of a quadrupole:

\begin{eqnarray}
\int_0^\infty \phi(r)d^3r
  &=&\int_0^{r_0}\left(\frac{Q}{4\pi\epsilon_0 r_0^3}
      - \frac{Q}{4\pi\epsilon_0 r_0^2 r}\right)4\pi r^2dr \\
  &=&\frac{Q}{\epsilon_0}\left[\frac{r^3}{3r_0^3}-\frac{r^2}{2r_0^2}\right]_0^{r_0}
  \\
  &=&-\frac{Q}{6\epsilon_0} \label{eqn:Qtoalpha}
\end{eqnarray}

In other words, a scalar quadrupole moment $Q$ will change the
integrated value of the electrostatic potential by an amount
$\frac{-Q}{6\epsilon_0}$. In the limit of $r_0 \rightarrow 0$ a point
quadrupole moment produces no field, but acts as a delta function
addition to the potential. If the average potential of a system is
fixed, and the Ewald summation fixes it to zero by ignoring the
$g=0$ Fourier components, then the addition of such a quadrupole
moment produces a shift of the potential throughout the cell.

This is analogous to the $\alpha$ term of a pseudopotential, which
represents how the integral of the local part of the pseudopotential
differs from that of the corresponding Coulomb potential. The ion-ion
interaction term in the total energy is accounted for in the Ewald
sum, which assumes that the ions are outside of each others'
pseudopotential core radii, and thus in the region where the
pseudopotentials are identical to Coulomb potentials. But if one is
setting the average potential to zero, then the $\alpha$ term produces
a shift in the potential outside the core radius, and this leads to
the `$Z\alpha$' energy term identified by Ihn\cite{Ihn79} and repeated here
as equation~\ref{eqn:NC}.

\section{The Makov-Payne correction for charged systems}

The basic expression for the energy of a system in Density Function
Theory is given by

\begin{equation}
E = E_k + E_{ee} + E_{Ewald} + E_{ie} + E_{XC} + \frac{|Z|}{V}\sum
\alpha_i \label{eqn:Etot}
\end{equation}

where $E_k$ is the kinetic energy of the electrons, $E_{ee}$ is the
Hartree energy, describing the Coulomb part of the electron-electron
interaction, $E_{Ewald}$ is the Ewald energy describing the ion-ion
Coulomb interaction, $E_{ie}$ is the Coulomb interaction between the
electrons and the ions,  $E_{XC}$ is the exchange-correlation energy.
Finally there is the `$Z\alpha$' term of equation~\ref{eqn:NC}. It arises
because each of $E_{ee}$, $E_{Ewald}$ and $E_{ie}$ is infinite, with
the infinities arising from the DC component of the potential in
Fourier space. In a neutral system these infinities cancel, and
numerically this cancellation is achieved by setting the average value
of the each potential to zero. The ionic potential is treated as
Coulombic in the Ewald energy, but as arising from the ions'
pseudopotentials in the $E_{ie}$ term. Given that an ion's
pseudopotential and Coulomb potential are identical outside of the
pseudopotential's core radius, and that the core radius should be
sufficiently small that no other ion lies within it, this would appear
not to matter. But because the integrated potential is set to zero,
and the integrals of the Coulomb and pseudopotentials differ, it leads to a
relative shift of the two potentials, which is then corrected by the
final `$Z\alpha$' term.

To this may be added corrections for the long-ranged effects of
dipoles and net charges. The energy correction for charged systems
proposed by Makov and Payne for a charged system in a cubic
cell\cite{MP95} is

\begin{equation}
- \frac{q^2M}{8\pi\epsilon_0L} -\frac{qQ}{6 \epsilon_0 V} \label{eqn:MP}
\end{equation}

where $q$ is the net charge, $M$ the Madelung constant of the
lattice, $Q$ the total quadrupole moment, $L$ the side-length of the
cube, and $V=L^3$ the cell volume. They demonstrated that this
correction removed terms decaying slower than $L^{-5}$, and they used
the total electronic charge in equation~\ref{eqn:NC}. The first term,
the Madelung energy, was well-known, but the second quadrupole term
was novel.

For a charged system, it matters whether the $Z$ in the final term of
equation~\ref{eqn:Etot} refers to the total electronic charge, as was
conventional when the Makov and Payne paper was written, or the total
ionic charge, as is conventional now. So to reproduce their work it is
necessary to consider the `$Z\alpha$' term in conjunction with the
quadrupole term from equation~\ref{eqn:MP}.

On adding their `$Z\alpha$' term to their quadrupole term one obtains

\begin{equation}
  E_{MP}=-\frac{qQ}{6 \epsilon_0 V} + \frac{Z-q}{V}\sum \alpha_i \label{eqn:MP_total}
\end{equation}

with $Z$ as the total ionic charge, and thus $Z-q$ the total electronic charge
as used in their `$Z\alpha$' term. This can be re-arranged to produce
the current `$Z\alpha$' term by writing

\begin{equation}
  E_{MP}=-\frac{q(Q+6\epsilon_0\sum \alpha_i)}{6 \epsilon_0 V} + \frac{Z}{V}\sum \alpha_i \label{eqn:MP_total2}
\end{equation}

Their total quadrupole moment, $Q$, was obtained by applying
equation~\ref{eqn:quad} to the valence charge density coupled with the
ions considered as point charges. Suppose that the
pseudopotentials themselves have some sort of intrinsic quadrupole
moment $Q_{ps}$ that should also be considered, so that in place of $Q$ one
should write $Q+\sum Q_{ps,i}$. If $Q_{ps,i}$ is defined to be
$6\epsilon_0\alpha_i$ then equation~\ref{eqn:MP_total2} follows immediately.

Thus the $Z$ in the $Z\alpha$ term has been restored to modern
convention of the total ionic charge by considering the spherical
quadrupole moment in the Makov-Payne correction to include an extra
term arising from the `quadrupole moments' of the pseudopotentials.

So one can either use the Makov-Payne correction with their definition
of the system's quadrupole moment and their use of the total
electronic charge in the $Z\alpha$ term, or equivalently one can
add the pseudopotential's `scalar quadrupole moments', defined as above
in terms of $\alpha$, to the system's quadrupole moment, and follow
the convention of using the total ionic charge in the $Z\alpha$ term.

It should be noted that this identification of the $\alpha$ term of a
pseudopotential with a quadrupole moment is not helpful in calculating
the scalar quadrupole moment of a system. To do that accurately one
needs the correct charge density within the pseudopotential's core
radius.

\section{Reconstructing a charge density}

An alternative approach might consider reconstructing the charge
density that would give rise to the pseudopotential. To do so using
just Gauss's Law, and thus ignoring the XC potential, is a very
artificial approach, but it does yield a useful result.

Pseudopotentials are spherically-symmetric, so the total charge within
a given radius, $q(r)$, is given by

\begin{eqnarray}
q(r) & = & \int_0^r \rho(r') 4\pi r'^2 dr' \label{eqn:qrho}
\end{eqnarray}

where $\rho$ is the charge density and $\phi$ the potential,
and with $q(r_c)$ being the total charge on the pseudopotential.

Gauss' Law is

\begin{align}
  \rho & =-\epsilon_0\nabla^2\phi \\
  & = -\epsilon_0 \frac{1}{r^2}\frac{\partial}{\partial r}\left(r^2
    \frac{\partial \phi}{\partial r} \right)
\end{align}

which can be inverted as

\begin{align}
\phi(r)&=\int_r^\infty \frac{1}{r'^2} \int_0^{r'}\frac{s^2 \rho(s)}{\epsilon_0} ds\, dr' \\
  &=\int_r^\infty \frac{1}{r'^2} \frac{q(r')}{4 \pi \epsilon_0} dr'
\end{align}

The scalar quadrupole moment of the charge distribution is defined as follows

\begin{align}
  Q & = \int_0^{r_c} r'^2 \rho(r') 4 \pi r'^2 dr' \\
    & = \left[ r^2 q(r) \right]^{r_c}_0-2 \int_0^{r_c} r q(r) dr \\
    & = r_c^2 q(r_c)-2\int_0^{r_c} r q(r) dr \label{eqn:Q} 
\end{align}

where integration by parts, and eqn~\ref{eqn:qrho}, move from the
standard definition to the final form.

And $\alpha$, the difference between the integral of the
pseudopotential, $\phi(r)$, and the integral of a Coulomb potential
from the same charge, is defined\cite{PayneTeter92,Martin04} as

\begin{align}
  \alpha & = \int_0^{r_c} \left( \frac{q(r_c)}{4\pi \epsilon_0 r'} - \phi(r') \right)
  4 \pi r'^2 dr' \\
  & = \frac{1}{\epsilon_0} \int_0^{r_c}  \left(  q(r_c) r' - r'^2 \int_{r'}^\infty \frac{q(s)}{s^2}ds 
  \right) dr'
\end{align}

The second part of the integral may be done by parts, and the $q(r_c)$ term
is integrated directly.

\begin{align}
  \alpha & = \frac{1}{\epsilon_0} \left( \frac{q(r_c) r_c^2}{2} - \left[ \frac{r'^3}{3}\int_{r'}^\infty \frac{q(s)}{s^2}ds \right]_0^{r_c} -\int_0^{r_c} \frac{r'^3}{3}\frac{q(r')}{r'^2} dr'
      \right) \\
    & = \frac{1}{\epsilon_0} \left( \frac{q(r_c) r_c^2}{2} - \left[ \frac{r'^3q(r_c)}{3}\int_{r'}^\infty \frac{1}{s^2}ds \right]_0^{r_c} -\frac{1}{3}\int_0^{r_c} r' q(r') dr' \right) \\
    & = \frac{1}{\epsilon_0} \left( \frac{q(r_c) r_c^2}{2}  -\left[ \frac{r'^3 q(r_c)}{3r'} \right]_0^{r_c} -\frac{1}{3}\int_0^{r_c} r' q(r') dr'  \right) \\
    & = \frac{1}{6\epsilon_0} \left(q(r_c) r_c^2- 2 \int_0^{r_c} r' q(r') dr' \right) \\
  & =  \frac{Q}{6\epsilon_0} \label{eqn:alpha_part}
\end{align}

Noting that $q(r)=q(r_c)$ for all $r \geq r_c$, and substituting from
equn~\ref{eqn:Q} for the last line.

So we conclude that
the scalar quadrupole moment of a charge distribution which, though
simple electrostatics, gives rise to a pseudopotential with a given
$\alpha$ term is $Q=6\epsilon_0 \alpha$. Whereas
equation~\ref{eqn:Qtoalpha} showed that a scalar quadrupole moment
gives this integrated potential, equation~\ref{eqn:alpha_part} gives
the more general result that $Q$ and $\alpha$ obey this relationship
for any potential arising from a spherically-symmetric charge density.

\section{The ionisation energy of M\lowercase{g}}

\begin{figure}
	\begin{center}
	\includegraphics[width=0.5\textwidth]{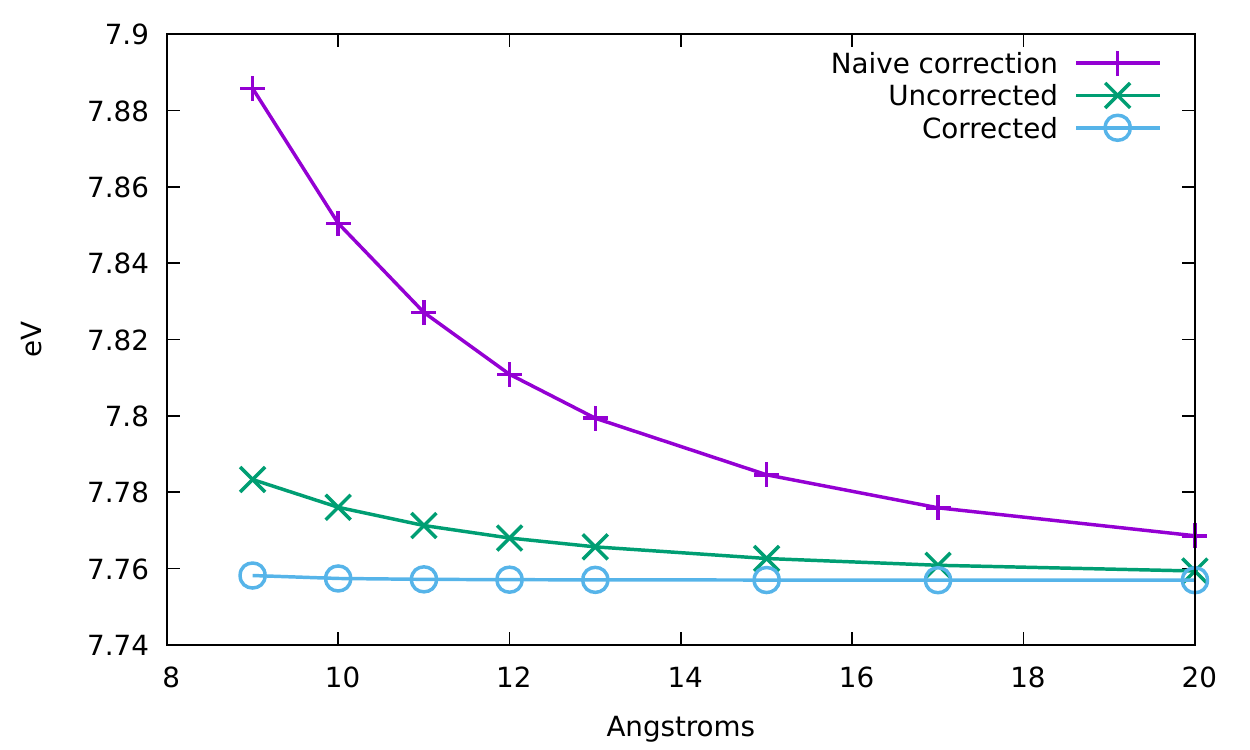}
	\end{center}
	\caption{The ionisation energy of magnesium, calculated in
          cubic boxes of increasing side length. All energies are
          shown after a simple Madelung correction. Those labelled
          `uncorrected' have no further correction, those labelled
          `naively corrected' also have the Makov-Payne correction
          naively applied to the output of a modern DFT code, and
          those labelled `corrected' have the  Makov-Payne correction
          correctly applied.}
	\label{fig:Mg}
\end{figure}

An example used by Makov and Payne in their paper introducing this
correction was the ionisation energy of Mg.  They calculated the
energy difference between Mg and Mg$^+$ in cubic boxes with sides
ranging from 9\AA{} to 13\AA.  These calculations are now repeated to
show the difference between using just a Madelung energy correction,
adding the quadrupole correction in the form described by their paper,
and adding it in the form described here with the quadrupole
moment including a $6\epsilon_0 \alpha$ term. A norm-conserving Mg$^{2+}$
pseudopotential and the LDA XC functional were used, as they would
have done. Castep\cite{CASTEP} was used, and c2x\cite{c2x} to
calculate the quadrupole moment in a post-processing step. Figure~\ref{fig:Mg}
repeats their figure 3(b), but a range of 9\AA{} to 20\AA{} is used.

The energies after the Madelung correction differ. Theirs converge
from below, whereas the results in this paper converge from
above. This is to be expected, as the different treatment of the $Z$
in the `$Z\alpha$' term means that these calculations are not
identical. This is the reason why the correction they proposed no
longer works in the precise form they gave. The experimental value of
the ionisation energy of Mg is 7.646eV\cite{Kaufman91}, and even this
simple calculation which yields 7.76eV is quite close, and closer than
Makov and Payne's 1995 result of around 7.95eV. Pseudopotential
improvements may account for the difference. Repeating the calculations
with a more modern XC functional, PW91\cite{PW91}, produces a result
of 7.66eV.

The calculated scalar quadrupole moment of the electron density in the
Mg$^+$ system is around $-2.45$e\AA{}$^2$. The norm-conserving
pseudopotential generated by Castep has an $\alpha$ corresponding to a
quadrupole moment of $+3.08$e\AA{}$^2$. Thus the electronic quadrupole
moment has the opposite sign to the total moment produced by including
the $6\epsilon_0 \alpha$ term from the pseudopotential, and a naive
attempt to use the quadrupole term from the Makov-Payne correction
results in a correction of the wrong sign which makes the error
worse. In this example, the unmodified correction uses a quadrupole
moment of $-2.45$e\AA{}$^2$ whereas it should use $+0.63$e\AA{}$^2$,
so the unmodified correction is almost four times too large, as well
as having the wrong sign.

\section{The ionisation energy of benzene}

\begin{figure}
	\begin{center}
	\includegraphics[width=0.5\textwidth]{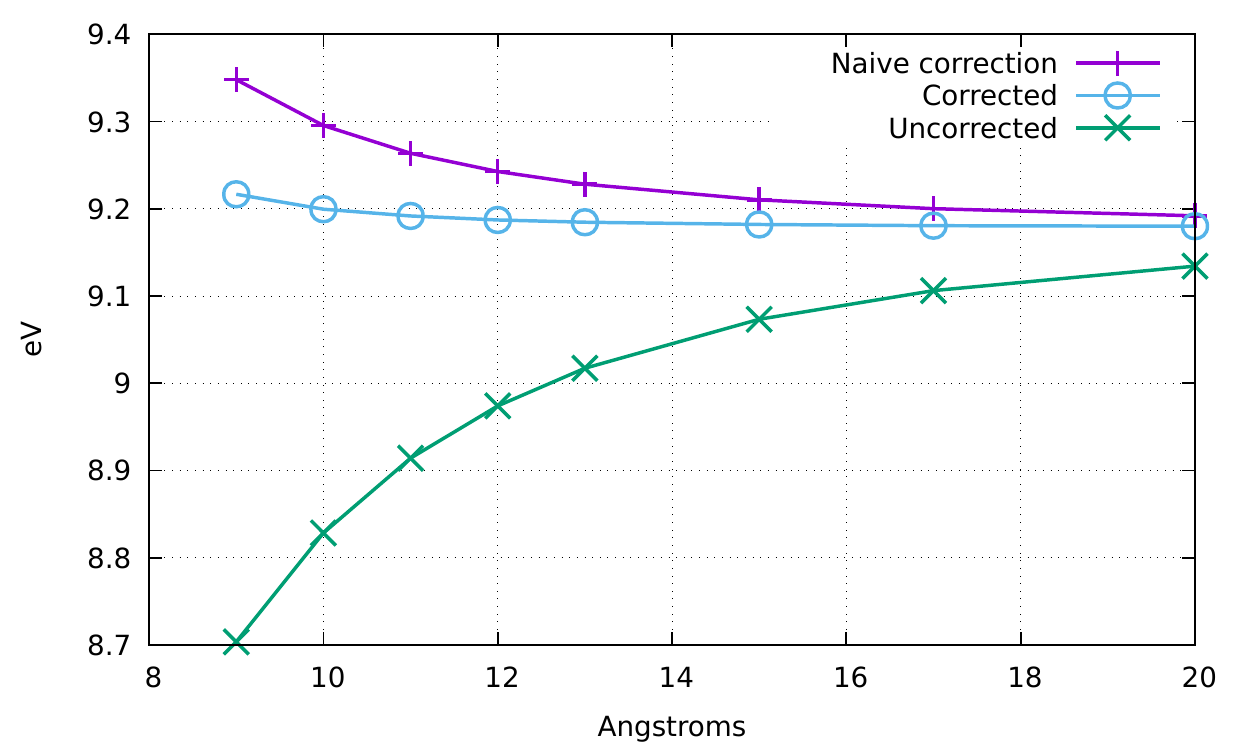}
	\end{center}
	\caption{The ionisation energy of benzene, calculated in cubic
          boxes of increasing side length, in the same manner as
          figure~1. All energies are shown after a simple Madelung
          correction. Two further curves show the result of applying the
          Makov-Payne correction naively to the output of a modern DFT
          code, and applying it correctly.}
	\label{fig:benzene}
\end{figure}

As a second example the ionisation of benzene is considered. The
benzene ion differs from Mg$^+$ in some important aspects. It is not
spherically symmetric, and it is much larger, with the charge
delocalised across the whole ion. Thus it fits less well with a
theory based on point multipole expansions. Relaxation of the atomic
positions was not performed.

The sum of the $\alpha$ terms of the pseudopotentials generated by
Castep corresponds to a quadrupole moment of $+3.2$e\AA{}$^2$, and
the scalar quadrupole moment of the ions considered as point charges,
plus the electron density, is around $-15.3$e\AA{}$^2$. So now a
naive attempt to use the Makov-Payne correction results in a
correction of the correct sign, but overestimates the magnitude by
about a quarter, for the correct total quadrupole moment
including the contribution from the pseudopotentials is around
$-12.1$e\AA{}$^2$, but ignoring the pseudopotential term gives rise to
a larger moment of $-15.3$e\AA{}$^2$, and hence an erroneously large
Makov-Payne correction.

Figure~\ref{fig:benzene} shows the result of applying the Makov-Payne
correction as a \textit{post hoc} energy correction both with, and
without, considering the contributions of the pseudopotentials to the
quadrupole moment. Although convergence for this more extended system
is slower than for the  Mg$^+$ ion, the correction is still very
helpful.

\begin{figure}
	\begin{center}
	\includegraphics[width=0.5\textwidth]{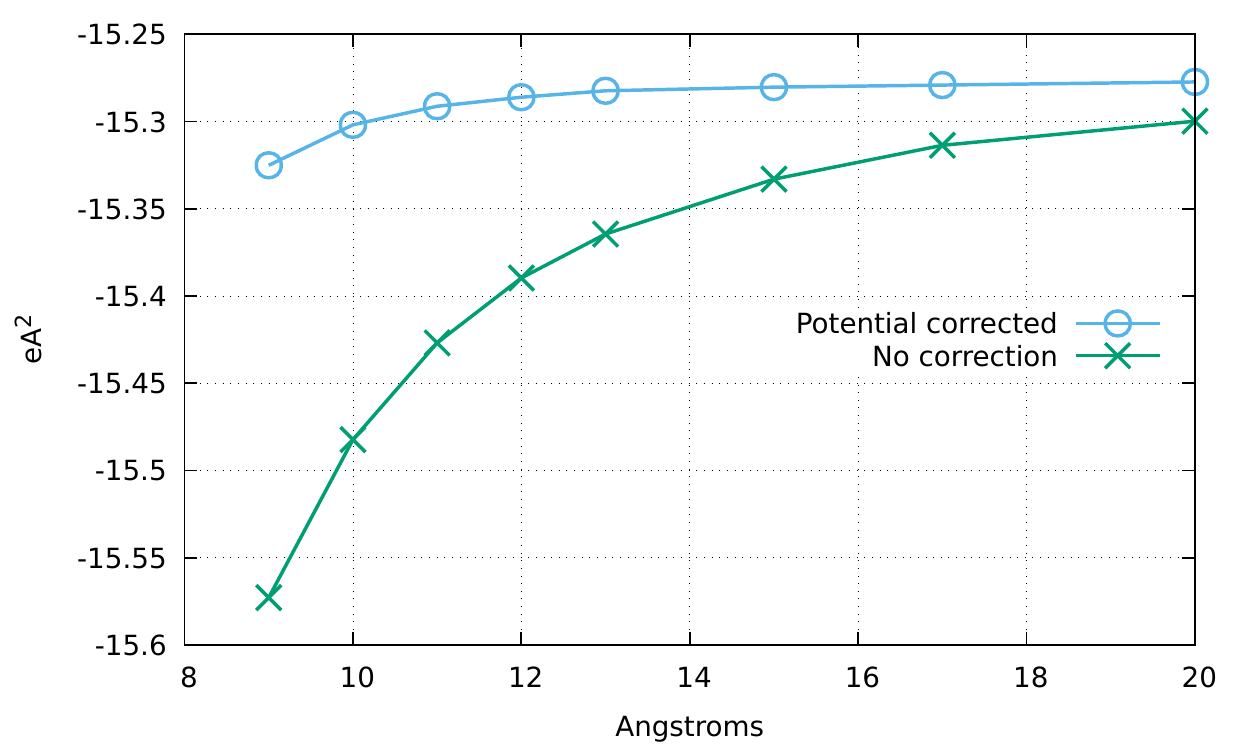}
	\end{center}
	\caption{The scalar quadrupole moment of the benzene ion,
          calculated in cubic boxes of increasing side length. Moments
          are show before and after applying a correcting quadratic
          potential to compensate for the potential arising from the
          uniform density of the background jellium.}
	\label{fig:benzene_Q}
\end{figure}

Ideally such corrections are applied self-consistently, rather than in
a \textit{post hoc} fashion. The extra energy term which depends on
the quadrupole moment ought to be considered as giving rise to a
potential which acts on the electrons so as to penalise configurations
which lead to energetically unfavourable moments. This was not done,
and the omission results in a quadrupole moment which varies slightly
with cell size. The moment excluding the pseudopotentials varied from
$-15.57$e\AA{}$^2$ in a 9\AA{} cell to $-15.27$e\AA{}$^2$ in the
20\AA{} cell.

Setting the zero frequency Fourier component of the charge density to
zero, as is necessary to prevent the energy per unit cell diverging,
is equivalent to introducing a uniform neutralising background charge,
often called `jellium.' This produces an unwanted potential. The
density of the jellium is $-q/V$ where $q$ is the net charge of the
cell and $V$ its volume. The potential, $\phi$, which arises can be
obtained from Gauss's Law if one assumes that the potential is
spherically symmetric. This assumption should be approximately valid
close to the centre of the cubic cell.

\begin{align}
  \nabla^2\phi & = \frac{q}{\epsilon_0V} \\
   \frac{1}{r^2}\frac{\partial}{\partial r}\left(r^2
  \frac{\partial \phi}{\partial r} \right) & = \frac{q}{\epsilon_0V} \\
  \phi & = \frac{qr^2}{6\epsilon_0V}
\end{align}

This result immediately gives the Makov-Payne energy correction, for
the energy of a quadrupole moment, $Q$, in this potential is
$qQ/6\epsilon_0V$, and this unwanted energy term needs subtracting
from the energy obtained when jellium is included, just as this
unwanted potential needs subtracting. This term is also given by Komsa
et al.\cite{Komsa12} in their equation 20.

The difference in the convergence of the energy between using a
\textit{post hoc} correction, and correcting the potential within the
calculation, is slight. In a cell of side 9\AA{}, the error in the
energy after the \textit{post hoc} correction is about 36meV, and with
the corrected potential this error is reduced by about 5meV.
The difference is more marked in the calculated quadrupole moment, and
this is shown in figure~\ref{fig:benzene_Q}. Without the correction,
the quadrupole moment converges with a leading error term inversely
proportional to volume, and with it the leading error term appears to
be inversely proportional to the square of the volume.

\section{Eigenvalues of benzene}

Not only should the total energy of an isolated charged system
converge rapidly with increasing simulation cell size, and also its
charge density and the moments thereof, but so too should the
eigenvalues of the electronic bands. The \textit{post hoc} energy
correction of Makov and Payne makes no attempt to adjust the
eigenvalues, but they can be corrected by making a corresponding
correction to the potential. As given by  Komsa
\textit{et al.}\cite{Komsa12}, this is

\begin{equation}
\phi_{MP}=\frac{q\alpha}{4\pi\epsilon_0L}-\frac{Q}{6\epsilon_0V}-\frac{qr^2}{6\epsilon_0V} \label{eqn:K12}
\end{equation}

The same expression is also given by Dabo \textit{et
  al.}\cite{Dabo08,Dabo11}, save that they did not choose their origin
to set the dipole moment equal to zero.

The last of these terms was introduced in the previous section. The
other two are constants, so may be implemented as a \textit{post hoc}
shift to the eigenvalues. The middle term contains $Q$, and again the
quadrupole moment including the contributions from the
pseudopotentials should be used.

\begin{figure}
	\begin{center}
	\includegraphics[width=0.48\textwidth]{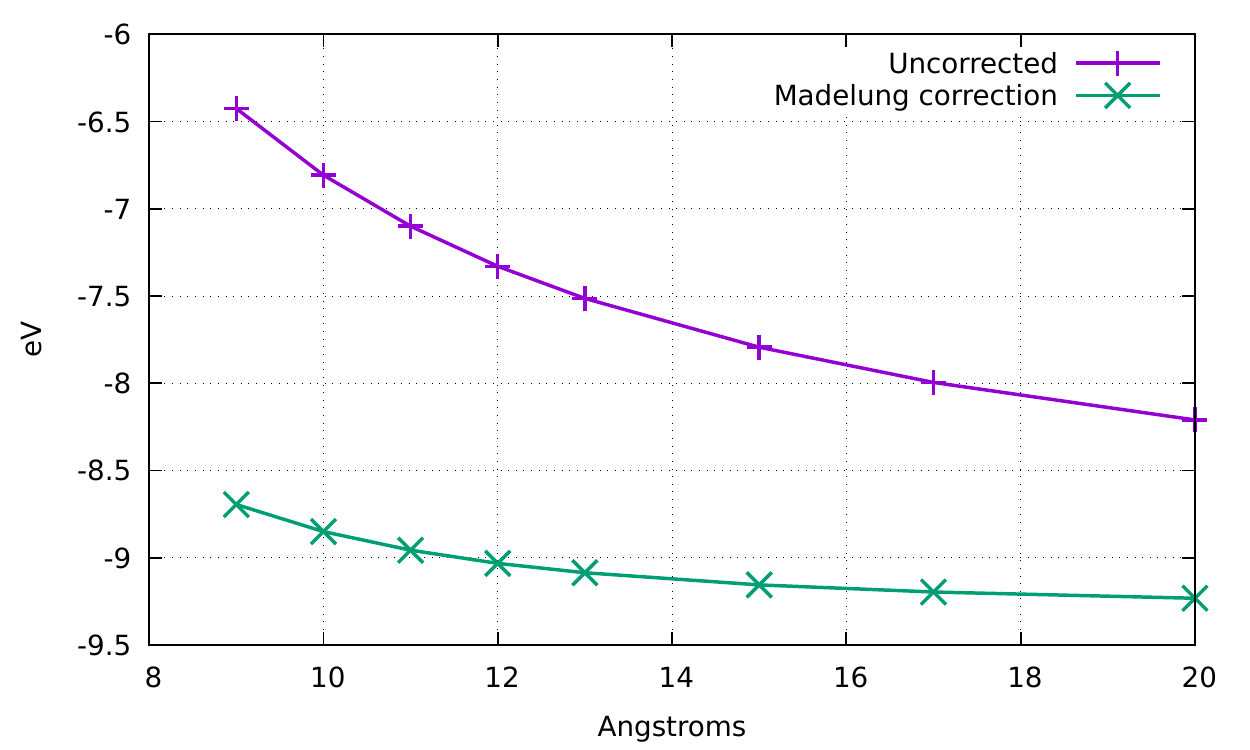}\hspace{0.02\textwidth}%
	\includegraphics[width=0.48\textwidth]{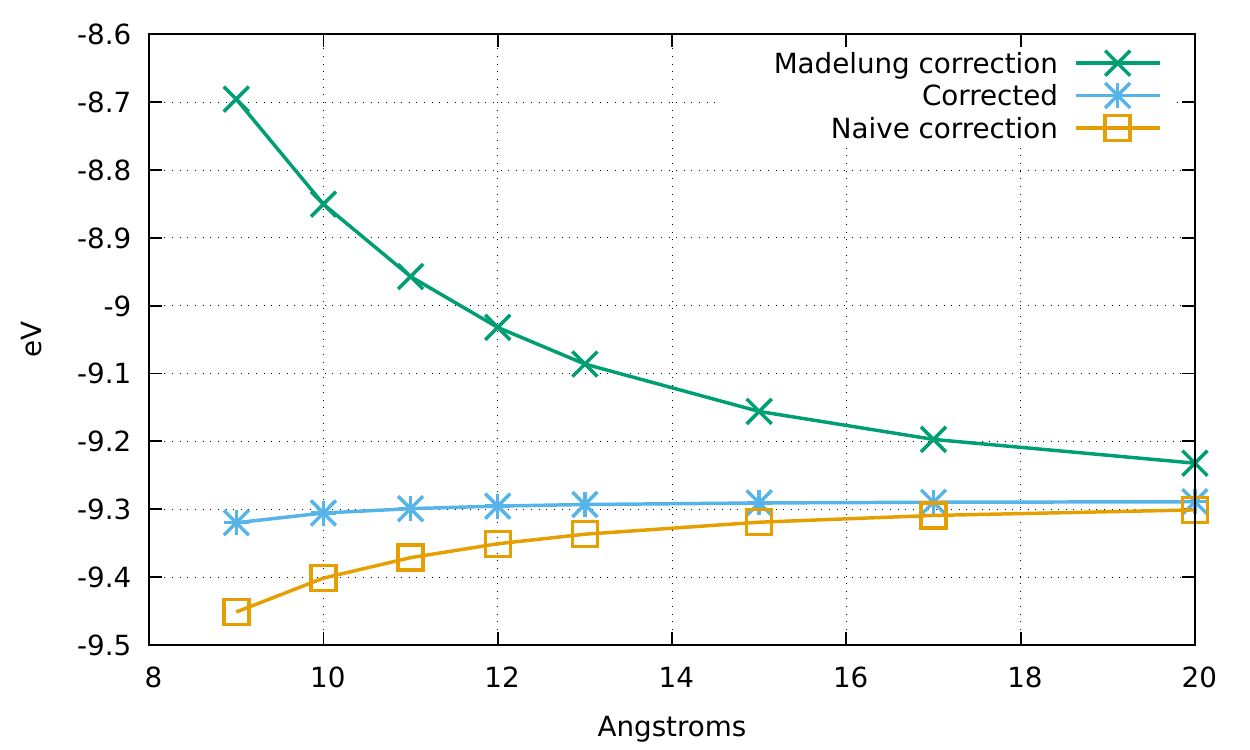}
	\end{center}
	\caption{The left-hand graph shows the convergence of the
          highest occupied eigenvalue of benzene with a $\frac{1}{2}$
          positive charge. It is shown with no corrections, and with
          the Madelung correction. The right-hand graph repeats the
          Madelung-corrected curve, but then adds the quadrupole
          correction with the pseudopotentials' `quadrupole moments'
          excluded, labelled `naive', and included, labelled `corrected'.}
	\label{fig:BE}
\end{figure}

As an illustration, the benzene ion is again chosen. The eigenvalue
considered is that of the highest (partially) occupied band when the
charge is $+\frac{1}{2}$, which is a value expected to approximate to
the ionisation energy.\cite{Slater72}

The left-hand part of figure~\ref{fig:BE} shows the slow
convergence of the eigenvalue when no corrections are applied, and the
much better convergence when simply the first Madelung constant shift
is added.

The right-hand part shows again the results after applying the
Madelung correction, but also after applying the second term of the
correction in a \textit{post hoc} fashion. To produce this figure, it
was assumed that the electronic and ion point charge contribution to
$Q$ was $-18.27$~e\AA$^2$, its value in the 20\AA{} cube, and the
pseuopotential contribution was $+3.18$~e\AA$^2$ so the correct value
for $Q$ was $-15.09$~e\AA$^2$.

With no correction the convergence is very slow, the leading term in
the error being of the form $1/L$. Adding the Madelung correction
greatly improves the convergence, with the leading error term becoming
$1/V$. Ignoring the pseudopotentials' moments when adding the next
correction term leads to over-correction, whereas including them gives
very good convergence. For these calculations the $r^2$ term of
equation~\ref{eqn:K12} was included in all four cases. The ionisation
energy predicted by this method of 9.29eV does not quite agree with the
9.18eV of the previous section, but this method produces estimates,
not exact results. For the purposes of this discussion, it is the
improved convergence with cell size which is of note.

\section{Conclusion}

Whilst the quadrupole correction for charged periodic systems proposed
by Makov and Payne appears to be incorrect when combined with more
recent treatments of the non-Coulomb pseudopotential energy term, it
can be corrected by assuming that the pseudopotentials themselves have
scalar quadrupole moments. Not only is this physically consistent with
the concept of a pseudopotential replacing a cloud of core electrons
which would have a quadrupole moment, but the required magnitude of
the moment can be obtained by an analytical calculation on a simple
model. To use the correction of Makov and Payne in codes which use the
total ionic charge in their `$Z\alpha$' energy term, a correction
equivalent to reverting to the total electronic charge in the
`$Z\alpha$' term is required, and this can be achieved by equating the
pseudopotential's $\alpha$ to a quadrupole moment. Failure to make
this adjustment can lead to the quadrupole term of the Makov-Payne
correction having not just the wrong magnitude, but also the wrong
sign and so increasing the error.

This observation is extended to the corrections to the potential
implied by the Makov and Payne correction. Including a
contribution from the pseudopotenials in the calculation of the
quadrupole moment greatly improves the convergence of the
eigenvalues.

\section{Acknowledgements}

The author acknowledges support from EPSRC grants numbers
EP/P034616/1 and EP/V062654/1.


\end{document}